\documentclass[12pt,aps,pra,groupedaddress,amssymb,amsfonts,nofootinbib,tightenlines]{revtex4} 

\usepackage{times,color}
\definecolor{purple}{rgb}{0.625,0.125,0.9375}

\usepackage{ekqc_b}

\ignore{
rm -R /tmp/ndgbnds; mkdir /tmp/ndgbnds
cp `texfls ndgbnds.log | perl -e '$a=<>; $a =~ s:/\S*(revtex4|natbib)\S*(\s|$)::g; print "$a\n"'` /tmp/ndgbnds
perl -i -n -e 's/^

cp /usr/share/texmf/tex/latex/tools/calc.sty /tmp/ndgbnds/phcalc.sty
perl -i -n -e 's/ProvidesPackage\{calc\}/ProvidesPackage\{phcalc\}/; print;' /tmp/ndgbnds/phcalc.sty
perl -i -n -e 's/usepackage\{calc\}/usepackage\{phcalc\}/; print;' /tmp/ndgbnds/ekqc_b.sty

(cd /tmp/ndgbnds; tar czvf ndgbnds.tar.gz *)
}

\ignore{
rm -R /tmp/ndgbnds; mkdir /tmp/ndgbnds
cp `texfls ndgbnds.log | perl -e '$a=<>; $a =~ s:/\S*(revtex4|natbib)\S*(\s|$)::g; print "$a\n"'` /tmp/ndgbnds
perl -i -n -e 's/^

cp /usr/share/texmf/tex/latex/tools/calc.sty /tmp/ndgbnds/phcalc.sty
perl -i -n -e 's/ProvidesPackage\{calc\}/ProvidesPackage\{phcalc\}/; print;' /tmp/ndgbnds/phcalc.sty
perl -i -n -e 's/usepackage\{calc\}/usepackage\{phcalc\}/; print;' /tmp/ndgbnds/ekqc_b.sty

cp /usr/share/texmf/tex/latex/base/ifthen.sty /tmp/ndgbnds/phifthen.sty
perl -i -n -e 's/ProvidesPackage\{ifthen\}/ProvidesPackage\{phifthen\}/; print;' /tmp/ndgbnds/phifthen.sty
perl -i -n -e 's/usepackage\{ifthen\}/usepackage\{phifthen\}/; print;' /tmp/ndgbnds/ekqc_b.sty

(cd /tmp/ndgbnds; tar czvf ndgbnds.tar.gz *)
}

\begin{document}
\title{Bounds on the Probability of Success of Postselected Non-linear Sign Shifts Implemented with Linear Optics}
\author{E. Knill}
\email[]{knill@lanl.gov}
\affiliation{Los Alamos National Laboratory, MS B256, Los Alamos, NM 87545} 

\date{\today}

\begin{abstract}
The fundamental gates of linear optics quantum computation are
realized by using single photons sources, linear optics and
photon counters. Success of these gates is conditioned on the pattern
of photons detected without using feedback.  Here it is shown that the
maximum probability of success of these gates is typically strictly
less than $1$.  For the one-mode non-linear sign shift, the
probability of success is bounded by $1/2$. For the conditional sign
shift of two modes, this probability is bounded by $3/4$.
These bounds are still substantially larger than the highest
probabilities shown to be achievable so far, which are $1/4$ and
$2/27$, respectively.
\end{abstract}


\maketitle

\section{Introduction}

It has recently been shown that it is possible, in principle, to
scalably quantum compute with single photon sources, linear optics and
photon counters~\cite{knill:qc2000e}\footnote{See
also~\cite{gottesman:qc2000a} for an alternative approach
and~\cite{yoran:qc2003a} for a significant improvement}. Key elements
in the scheme that makes this possible are optical gates that use
helper photons, linear optics and postselection on specific photon
counts to realize simple non-linear operations on one or more
modes. Two such gates are the one-mode non-linear sign shift
\begin{equation}
\textsf{NS}:\alpha\ket{0}+\beta\ket{1}+\gamma\ket{2}\rightarrow
            \alpha\ket{0}+\beta\ket{1}-\gamma\ket{2}
\end{equation}
and the two-mode conditional sign shift 
\begin{equation}
\textsf{CS}:\alpha\ket{00}+\beta\ket{10}+\gamma\ket{01}+\delta\ket{11}
            \rightarrow
            \alpha\ket{00}+\beta\ket{10}+\gamma\ket{01}-\delta\ket{11}.
\end{equation}
Here $\ket{j}$ is the state with $j$ photons in one mode and
$\ket{jk}$ is the state with $j$ photons in the first and $k$ photons
in the second mode. How these gates act on states other than those
explicitly given does not matter for current purposes.  To
efficiently use these gates, one would like to implement them with as
high a probability of success as possible. To do so one may use single
helper photons in helper modes, apply a linear optics transformation
(that is, a series of beam splitters and phase shifters), and a
combination of photon counting measurements of the helper modes.  In
the remainder of this report, a procedure using single helper photons
and linear optics is called an LOP procedure. LOP states are those
obtained by an LOP procedure from the vacuum. Postselection based on
measured photon counts is abbreviated as PC. All procedures considered
here are assumed not to involve feedback from PC, that is, they consist
of LOP followed by PC.  Currently, the highest probabilities of success
achieved for implementing $\textsf{NS}$ and $\textsf{CS}$ with LOP
followed by PC are $1/4$~\cite{knill:qc2000e} and
$2/27$~\cite{knill:qc2001g}, respectively.  What are the maximum
probabilities of success $P_{\mathrm{max}}(\textsf{NS})$ and
$P_{\mathrm{max}}(\textsf{CS})$ for realizing these gates with LOP
followed by PC?  In~\cite{knill:qc2001g} it was shown that these
probabilities cannot be exactly one. The main result of this report is
to show that $P_{\mathrm{max}}(\textsf{NS})\leq 1/2$ and
$P_{\mathrm{max}}(\textsf{CS})\leq 3/4$.  To prove these bounds, the
gates are used to create special two-photon states. The next step is
to obtain upper bounds on the maximum overlaps of these states with
states that can be generated with LOP. Since high probability of
success for the gates implies high overlap with the state obtained
just before postselection, the desired bounds can be obtained. The
bounds on the overlaps are derived by considering photon statistics of
LOP states. The techniques can be applied to obtain bounds on the
probability of success of other postselected gates.

\section{Upper Bounds: \textsf{NS}}

To bound $P_{\mathrm{max}}(\textsf{NS})$ from above, assume that we
can implement $\textsf{NS}$ using LOP followed by PC with probability
of success $p$.  The following procedure creates the two photon state
from single photon states with probability of success $p$:
\begin{itemize}
\item[1.] Prepare the state
$\kets{11}{ab}=\crtops{a}\crtops{b}\vacuum$ consisting of one photon
in each of modes $\sysfnt{a}$ and $\sysfnt{b}$.
Here $\vacuum$ is the vacuum state and $\crtops{x}$ is the creation
operator for mode $\sysfnt{x}$.
\item[2.] Set $\alpha=\cos(\pi/8)$ and $\beta=\sin(\pi/8)$.
Use the beam splitter that transforms
$\kets{10}{ab}\rightarrow\alpha\kets{10}{ab}+\beta\kets{01}{ab}$
and $\kets{01}{ab}\rightarrow-\beta\kets{10}{ab}+\alpha\kets{01}{ab}$.
Writing $U$ for the unitary operator implemented by this beam splitter,
$U$'s action can be derived from how it transforms the
annihilation and creation operators for the modes.
That is, $U\crtops{a}U^\dagger = \alpha\crtops{a}+\beta\crtops{b}$
and $U\crtops{b} U^\dagger = -\beta\crtops{a}+\alpha\crtops{b}$,
where $\killops{l}$ and $\crtops{l}$ are the annihilation and creation
operators for mode $\sysfnt{l}$, respectively. 
The following state is obtained after applying this beam splitter:
\begin{eqnarray}
U\kets{11}{ab} &=&
  U\crtops{a}\crtops{b}\vacuum \\
  &=&
   U\crtops{a}U^\dagger U\crtops{b}U^\dagger U\vacuum\\
  &=&
(\alpha\crtops{a}+\beta\crtops{b})(-\beta\crtops{a}+\alpha\crtops{b})\vacuum\\
 &=& (-\alpha\beta\crtops{a}^2+(\alpha^2-\beta^2)\crtops{a}\crtops{b}
        +\alpha\beta\crtops{b}^2)\vacuum\\
 &=& -\sqrt{2}\alpha\beta\kets{20}{ab}+(\alpha^2-\beta^2)\kets{11}{ab}
        +\sqrt{2}\alpha\beta\kets{02}.
\end{eqnarray}
\item[3.] Apply $\textsf{NS}$ to mode $\sysfnt{a}$ to obtain
\begin{eqnarray}
\sqrt{2}\alpha\beta\kets{20}{ab}+(\alpha^2-\beta^2)\kets{11}{ab}
        +\sqrt{2}\alpha\beta\kets{02}{ab}\hspace*{-2in}\nonumber\\
  &=& {1\over 2}(\sin(\pi/4)\crtops{a}^2 + 2\cos(\pi/4)\crtops{a}\crtops{b}
         +\sin(\pi/4)\crtops{b}^2)\vacuum\\
  &=& {1\over \sqrt{2}}({1\over \sqrt{2}}(\crtops{a}+\crtops{b}))^2\vacuum
\end{eqnarray}
with probability of success $p$. Here, $\vacuum$ is the vacuum state,
that is, the state with no photons in any of the modes under consideration.
\item[4.] By using a $50/50$ beam splitter that maps
${1\over\sqrt{2}}(\kets{10}{ab}+\kets{01}{ab})\rightarrow\kets{10}{ab}$,
the state ${1\over\sqrt{2}}\crtops{a}^2\vacuum = \kets{20}{ab}$ is
obtained.
\end{itemize}
The effect of the above procedure is unchanged if PC is delayed until
the end. Let $\rho$ be the final state (density matrix) on mode
$\sysfnt{a}$ just before postconditioning.  Because postconditioning
on a measurement of modes other than $\sysfnt{a}$ to obtain
$\ketbras{2}{2}{a}$ is possible and the probability of success is $p$,
$\rho$ can be expressed as a mixture $p\ketbras{2}{2}+(1-p)\rho'$ for
some state $\rho'$. To bound $p$ from above requires the following
result:

\begin{Theorem}
\label{thm:expected_n}
Let $\varrho$ be an LOP state.  Then $\varrho$'s expected number of
photons in any mode is at most $1$.
\end{Theorem}

The expected number of photons in mode $\sysfnt{a}$ for $\rho$ is
given by $2p+x$, where $x\geq 0$. It follows that $p\leq 1/2$,
establishing the desired bound on $P_{\mathrm{max}}(\textsf{NS})$.

\proofoft{Theorem}{\ref{thm:expected_n}}
Let the initial state before applying the linear optics transformation
be given by
\begin{equation}
\ket{\psi}=\crtops{1}\ldots\crtops{k}\vacuum,
\end{equation}
where $k$ is the number of single photons used.  Let the linear optics
transformation $U$ act on modes $\sysfnt{1}$ through $\sysfnt{n}$,
$\sysfnt{n}\geq \sysfnt{k}$.  The transformation is completely
determined by its $n\times n$ unitary matrix $\hat
U=(u_{\sysfnt{jl}})$ determined by $U^\dagger\crtops{l} U =
\sum_{\sysfnt{j}}u_{\sysfnt{jl}}\crtops{j}$~\cite{reck:qc1994a}.
Without loss of generality, consider
the expected number of photons in the first mode after $U$ has been
applied. Compute
\begin{eqnarray}
\langle\numberops{1}\rangle
  &=& \bra{\psi}U^\dagger\crtops{1}\killops{1}U\ket{\psi}\\
  &=& \bra{\psi}U^\dagger\crtops{1}UU^\dagger\killops{1}U\ket{\psi}\\
  &=& \bra{\psi}(\sum_{\sysfnt{j}}u_{\sysfnt{j1}}\crtops{j})
                (\sum_{\sysfnt{l}}\bar u_{\sysfnt{l1}}\killops{l})\ket{\psi}\\
  &=& \sum_{\sysfnt{jl}}u_{\sysfnt{j1}}u_{\sysfnt{l1}}
         \bra{\psi}\crtops{j}\killops{l}\ket{\psi}\\
  &=& \sum_{\sysfnt{j}=1}^{\sysfnt{k}}|u_{\sysfnt{j1}}|^2\label{eq:n1}\\
  &\leq& 1.
\end{eqnarray}
The second last step follows because $\ket{\psi}$ has well defined
photon numbers in each mode, with none in modes beyond mode
$\sysfnt{k}$. The last step follows by unitarity of $\hat U$.  
\qed

\section{Upper Bounds: \textsf{CS}}

The bound on $P_{\mathrm{max}}(\textsf{CS})$ is obtained in the same
way as that on $P_{\mathrm{max}}(\textsf{NS})$.  Assume that we can
implement $\textsf{CS}$ with probability of success $p$. The first
step is to show that one can create a state with expected number of photons
$4/3$ in a mode using one instance of $\textsf{C}$.
\begin{itemize}
\item[1.] Prepare the state $\kets{110}{abc}$.
\item[2.] Use a beam splitter on modes $\sysfnt{b}$ and $\sysfnt{c}$
to make the state 
\begin{equation}
{1\over\sqrt{3}}\kets{110}{abc}+
    {\sqrt{2}\over\sqrt{3}}\kets{101}{abc}.
\end{equation}
\item[3.] Use a beam splitter on modes $\sysfnt{a}$ and
$\sysfnt{b}$ that transforms
$U_1\ket{10}=\cos(\pi/8)\kets{10}{ab}
             -\sin(\pi/8)\kets{01}{ab}$
and $U_1\ket{01}= \sin(\pi/8)\kets{10}{ab}
                  +\cos(\pi/8)\kets{01}{ab}$.
This gives
\begin{equation}
{1\over\sqrt{3}}U_1\kets{110}{abc}
  +{\sqrt{2}\over\sqrt{3}}\left(\cos(\pi/8)\kets{101}{abc}
                           -\sin(\pi/8)\kets{011}{abc}\right).
\end{equation}
\item[4.] Apply $\textsf{CS}$ on to modes $\sysfnt{b}$ and $\sysfnt{c}$
with probability of success $p$ to obtain
\begin{equation}
{1\over\sqrt{3}}U_1\kets{110}{abc}
  +{\sqrt{2}\over\sqrt{3}}\left(\cos(\pi/8)\kets{101}{abc} +
                           \sin(\pi/8)\kets{011}{abc}\right).
\end{equation}
\item[5.]
Apply the inverse of the beam splitter used in step 3. The state 
is now
\begin{eqnarray}
\ket{\psi}&=&
  {1\over\sqrt{3}}\kets{110}{abc}+
  {\sqrt{2}\over\sqrt{3}}\Big(
     (\cos(\pi/8)^2-\sin(\pi/8)^2)\kets{101}{abc}
     \\&&\iboxlike{${1\over\sqrt{3}}\kets{110}{abc}+
  {\sqrt{2}\over\sqrt{3}}((\;\;$}+2\cos(\pi/8)\sin(\pi/8)\kets{011}{abc}\Big)\\
  &=& {1\over\sqrt{3}}(\kets{110}{abc}+\kets{101}{abc}+\kets{011}{abc}).
\end{eqnarray}
\end{itemize}
The claim is that the logical mode associated
with annihilation operator 
$\killops{l}={1\over\sqrt{3}}(\killops{a}+\killops{b}+\killops{c})$
has expected photon number $4/3$. This logical mode can be transformed
into mode $\sysfnt{a}$ by a linear optics transformation.
Using Thm.~\ref{thm:expected_n} we can conclude, as before,
that the maximum probability with which this state can be obtained
is $3/4$. To prove the claim compute
\begin{eqnarray}
\bra{\psi}\numberops{l}\ket{\psi} &=& 
  \bra{\psi}
  \crtops{l}\killops{l}
  \ket{\psi}\\
  &=&
  {1\over 3}\bra{\psi}(
     \killops{a}+\killops{b}+\killops{c})
     (
     \crtops{a}+\crtops{b}+\crtops{c})
  \ket{\psi} \\
  &=&
{1\over 3} \Bigg(
   {2\over\sqrt{3}}(\bras{100}{abc}+\bras{010}{abc}+\bras{001}{abc})\\
  &&\iboxlike{${1\over 3}\Bigg(\;\;$}
   {2\over\sqrt{3}}(\kets{100}{abc}+\kets{010}{abc}+\kets{001}{abc})
  \Bigg) \\
 &=& {4\over 3}.
\end{eqnarray}

\section{Discussion}
The above results reduce the bounds on the probabilities of success of
$\textsf{NS}$ and $\textsf{CS}$ using LOP followed by PC to values
strictly below $1$. However, the gap between the highest probability
of the known procedures and the bounds found is still large.  An
obvious reason that the bounds found here are probably not optimal is
that they are insensitive to the type of measurement device used to
implement the postselection. That is, it doesn't matter whether a
photon counter or any arbitrarily more powerful measurement device is
used, the bounds are still valid.  Nevertheless, better bounds on the
probabilities of success may be obtainable without using properties of
photon counters.  For example, it may be possible to obtain better
bounds by using the $\textsf{NS}$ and $\textsf{CS}$ gates one or more
times to obtain states that are further from LOP states.  Note that
the two photon state can be obtained with probability $1/2$ with LOP
followed by PC: Apply a 50/50 beam splitter to the state $\ket{11}$
and condition on measuring no photons in the second mode. Here is an
example of states that can be investigated: It is not hard to see that
with one application of $\textsf{CS}$ to a state obtained from
$\ket{11}$, one can make the entangled state
$\ket{1100}+\ket{0011}$. It is plausible that this state can be
obtained with at most probability $1/2$ using LOP followed by PC.  The
density matrix for the first two modes is
$\rho=\ketbra{00}{00}+\ketbra{11}{11}$. If we consider a state $\mu$
obtained from single photons with linear optics, is it true that the
maximum $p$ for which $\mu=p\rho+(1-p)\varrho$ with $\varrho$ a
density matrix is $p=1/2$?

Because of their application to scalable linear optics quantum
computation, the postselected gates $\textsf{NS}$ and $\textsf{CS}$
and their variations are being studied both experimentally and
theoretically by many researchers.  Experimental work preparing for
the implementation of these gates has been reported
in~\cite{pittman:qc2001b,pittman:qc2002a}.  Related gates and schemes
have been investigated
in~\cite{rudolph:qc2001a,ralph:qc2001a,franson:qc2002a}.
Postselection techniques for implementing operations such as the above
have been studied
in~\cite{clausen:qc2003a,lapaire:qc2003a,scheel:qc2003a}.  Related
bounds, originally motivated by the problem of realizing a complete
Bell-basis measurement can be found
in~\cite{lutkenhaus:qc1999a,calsamiglia:qc2001a,loock:qc2003a}.

\begin{acknowledgments}
Many thanks to Leonid Gurvits for stimulating discussions. This work
was supported by the DOE (contract W-7405-ENG-36) and the NSA.
\end{acknowledgments}

\bibliography{journalDefs,qc}

\end{document}